%% file: main.tex
\documentclass[conference]{IEEEtran}
\IEEEoverridecommandlockouts
\usepackage{amsmath,amssymb,amsfonts}
\usepackage{algorithmic}
\usepackage[style=ieee]{biblatex}
\usepackage{booktabs}
\usepackage{csquotes}
\usepackage{enumitem}
\usepackage{graphicx}
\usepackage{siunitx}
\usepackage{textcomp}
\usepackage{threeparttable}
\usepackage{xcolor}
\def\BibTeX{{\rm B\kern-.05em{\sc i\kern-.025em b}\kern-.08em
    T\kern-.1667em\lower.7ex\hbox{E}\kern-.125emX}}

\usepackage{fancyhdr}
\begin{document}

\title{Towards automatic estimation of conversation floors within F-formations
\thanks{This research was partially funded by the Netherlands Organization for
Scientific Research (NWO) under project number 639.022.606.}
}

\author{\IEEEauthorblockN{Chirag Raman}
\IEEEauthorblockA{\textit{Delft University of Technology}\\
c.a.raman@tudelft.nl}
\and
\IEEEauthorblockN{Hayley Hung}
\IEEEauthorblockA{\textit{Delft University of Technology}\\
h.hung@tudelft.nl}
}

\maketitle
\thispagestyle{fancy}

\begin{abstract}
The detection of free-standing conversing groups has received significant
attention in recent years. In the absence of a formal
definition, most studies operationalize the notion of a conversation group
either through a spatial or a temporal lens. Spatially, the most commonly
used representation is the \textit{F-formation}, defined by social scientists as
the configuration in which people arrange themselves to sustain an interaction.
However, the use of this representation is often accompanied with the
simplifying assumption that a single conversation occurs within an F-formation.
Temporally, various categories have been used to organize conversational units;
these include, among others, \textit{turn}, \textit{topic}, and \textit{floor}.
Some of these concepts are hard to define objectively by themselves.
The present work constitutes an initial exploration into unifying these
perspectives by primarily posing the question: can we use the observation of
simultaneous speaker turns to infer whether multiple conversation floors exist
within an F-formation? We motivate a metric for the existence of distinct
conversation floors based on simultaneous speaker turns, and provide an analysis
using this metric to characterize conversations across F-formations of varying
cardinality. We contribute two key findings: firstly, at the average speaking
turn duration of about two seconds for humans, there is evidence for the
existence of multiple floors within an F-formation; and secondly, an increase
in the cardinality of an F-formation correlates with a decrease in duration of
simultaneous speaking turns.
\end{abstract}


\input{Sections/Introduction}

\input{Sections/Background}
\input{Sections/RelatedWork}
\input{Sections/Methodology}

\input{Sections/Dataset}
\input{Sections/Experiments}

\input{Sections/Conclusion}

\section*{Acknowledgment}

Chirag Raman thanks Stavros Makrodimitris, Madhumita Sushil, Giovanni Cassani,
Erik B. van den Akker, and  Yeshwanth Napolean for their time and thoughtfulness.

\renewcommand*{\bibfont}{\footnotesize}
\printbibliography

\end{document}

%% file: Sections/Introduction.tex
\section{Introduction}

Imagine a social scenario like a mingling or networking event. Interactions in
such a setting involve multiple dynamic conversations which are a medley of ever
evolving topics and partners. And yet, humans can instinctively navigate the
complexities of such encounters. How do we do this? We regulate our exchanges
both spatially and temporally using implicit social norms or explicit behavioural
signals \cite{vinciarelli_social_2009}. Furthermore, these cues could be either
verbal or non-verbal, expressed visually, vocally, or verbally through spoken language.

A deeper understanding of these group dynamics constitutes a natural objective
towards the realisation of machines with social skills. For instance, consider
a social robot approaching a group of people in a public space, or the use-case
of evaluating attendee experience at a conference poster session. In these and
other cases, having an understanding of the dynamics, and where channels of
social influence lie, would enable the artificial agent to develop increasingly
sophisticated policies for interaction or inference. Conversation groups have
been of importance in the application domains of social robotics
\cite{huettenrauch_investigating_2006, vazquez_parallel_2015, vazquez_maintaining_2016, vazquez_towards_2017},
activity recognition \cite{hutchison_unified_2012, tran_activity_2014},
social surveillance \cite{cristani_social_2011, farenzena_social_2013, bazzani_joint_2015},
and social signal processing \cite{groh_detecting_2010, hung_detecting_2011}.

\begin{figure}[t]
\makebox[\linewidth][c]{\includegraphics[width=\linewidth]{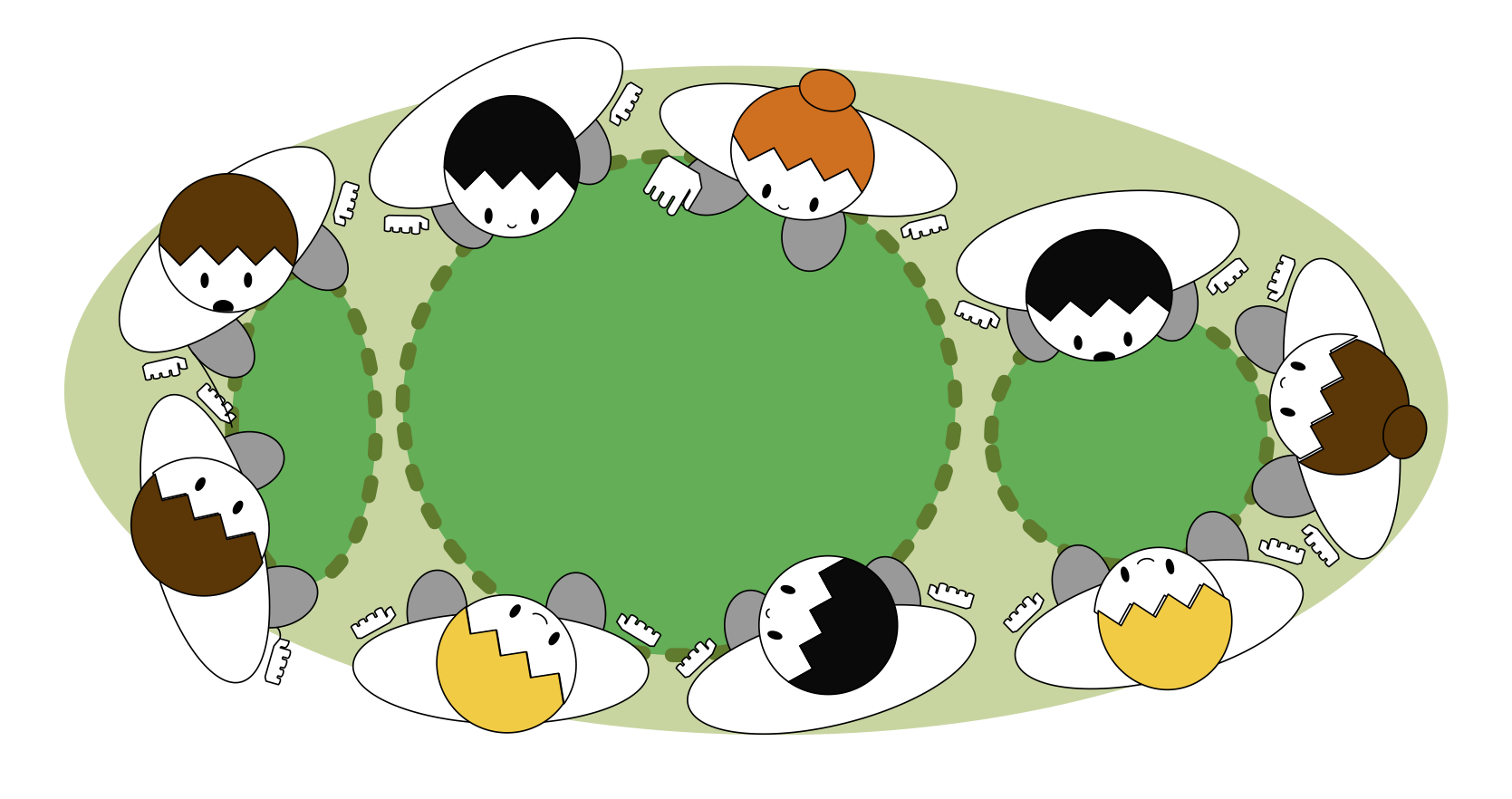}}
\caption{Depiction of a single F-formation with multiple conversation floors.
The darker green regions within dotted lines represent distinct simultaneous
conversation floors. Most works representing a conversing group as an F-formation
make the simplifying assumption that a single conversation occurs within an F-formation
with a joint focus of attention for all members. }
\label{fig:concept}
\end{figure}

Fundamental to the study of such conversations is defining the notion of a
free-standing conversational group (FCG). While it is easier to objectively
conceptualize an FCG in spatial terms in a scene of multiple interacting groups,
delineating the boundary of conversations poses a greater technical challenge.
We could think of separating conversations on the basis of topics, but this is
challenging if audio data is unavailable due to privacy concerns. We could
operationalize a conversation as a set of participating members, but this
membership is challenging to infer visually for non-speaking participants.
This often leads to the simplifying assumption in some literature that the focus
of an FCG is a single conversation. As we illustrate in Fig.~\ref{fig:concept},
and discuss in the following sections, this may not always be the case.

In the present work, we dive beyond the geometric bounds of an FCG to gain
a deeper understanding of the conversations occuring within it. In this initial
approach, we focus specifically on speaking participants as the most decisive
indicator of the existence of a conversation. Concretely, we pose the following
broad research questions:

\begin{enumerate}[label={RQ \arabic*.}, wide=0pt, leftmargin=*]
    \item Can we use observed speaker turns to infer the conversation floors
    within an F-formation?
    \item How does the cardinality of an F-formation affect the
    conversation floors developed within it?
\end{enumerate}

The ground truth for speaker turns in this work comes from manual annotations
of video data, mimicking use-cases where audio data might be unavailable due
to privacy concerns. Concretely, our contributions are as follows: conceptually,
we provide an indicator of distinct conversation floors that uses speaking turns
alone, and situate this indicator in schisming literature
\cite{sacks_simplest_1974, goodwin_forgetfulness_1987, egbert_schisming:_1997};
analytically, we provide evidence that multiple conversation floors exist within
an F-formation, and show that the cardinality of an F-formation correlates
negatively with turn duration of simultaneous speakers.

The rest of this paper is organized as follows. We describe some of the spatial
and temporal perspectives used to study FCGs in Section~\ref{sec:Background}.
In Section~\ref{sec:RelatedWork} we provide a review of literature involving the
use of these spatial or temporal notions, motivating the need to consider both of
these aspects in unison. In Section~\ref{sec:Methodology}, we propose an
operationalization of an indicator of distinct conversation floors,
building upon the concepts of conversation schisming. The dataset we use and
the experiments performed for answering the research questions are described in
Section~\ref{sec:Dataset} and Section~\ref{sec:Experiments} respectively.
Finally, Section~\ref{sec:Conclusion} summarizes our findings and concludes the paper.

%% file: Sections/Background.tex
\section{Background} \label{sec:Background}

\noindent \textbf{Spatial Factors.} One of the most common proxemic notions to
describe an FCG is Adam Kendon's \textit{Facing Formation}, or \textit{F-formation},
originally defined as:

\begin{displayquote}
An F-formation arises whenever two or more people sustain a spatial and
orientational relationship in which the space between them is one to which they
have equal, direct, and exclusive access. \parencite[p. 210]{kendon_conducting_1990}
\end{displayquote}

Kendon argues that activity is always located, and denotes the space in front
of a person that is used for the activity as the person's \textit{transactional segment}.
When two or more people come together to perform some activity, they are liable
to arrange themselves such that their individual transactional segments overlap
to create a joint transactional space. This joint space between the interactants
is called an \textit{o-space}. As we discuss in the next section,
many computational works involving the automatic detection of FCGs from video
focus on the detection of F-formations, often assuming that the transaction
involves a single conversation.

\hfill \break
\noindent \textbf{Temporal Factors.} The conversation of focus in an FCG, however,
is dynamic in nature. If conversations change over time, what are the temporal
units that describe their underlying structure? Some of the terms used in early
literature to organize conversational units include \textit{turn}, \textit{topic},
\textit{gap}, and \textit{floor}. \citeauthor{edelsky_whos_1981} provides an
excellent review of these concepts in \cite{edelsky_whos_1981}, stating that
most of these units were defined on the basis of some technical or mechanical
structure such as signals of speakers or auditors, ignoring the intention of
the participant. Using inferred participants' meanings rather than technical
definitions, \citeauthor{edelsky_whos_1981} defines turns and floors as follows:

\begin{displayquote}
The floor is defined as the acknowledged what's-going-on within a psychological
time/space. What's going on can be the development of a topic or a function
(teasing, soliciting a response, etc.) or an interaction of the two. It can be
developed or controlled by one person at a time or by several simultaneously or
in quick succession.\parencite[p. 405]{edelsky_whos_1981}
\end{displayquote}

%% file: Sections/RelatedWork.tex
\section{Related Work} \label{sec:RelatedWork}
\noindent \textbf{Detecting Conversational Groups.} In most works, a
conversational group is operationalized as an F-formation. Early work on the task
of detecting FCGs in video data developed concurrently from
two perspectives: those that estimate the location of the o-space using a
Hough-voting strategy \cite{cristani_social_2011, setti_multi-scale_2013}; or
those that view an F-formation as a set with individuals being assigned exclusive
membership \cite{hung_detecting_2011, setti_f-formation_2015}. There has also
been considerable work focused on incorporating temporal information for the
same task of detecting conversational groups
\cite{alameda-pineda_analyzing_2015, tran_activity_2014, vascon_detecting_2016, ricci_uncovering_2015}.
Notably, these approaches utilise the head pose as a proxy for Visual Focus of
Attention (VFoA) \cite{farenzena_social_2013} in addition to the body pose to
model F-formation membership, and assume a single conversation within an F-formation.
The assumption that members in a group have a single joint focus of attention is
seen in other works as well. \citeauthor{hung_investigating_2008} \cite{hung_investigating_2008}
model a single joint focus of visual attention of participants to estimate dominance
in groups.\citeauthor{vazquez_maintaining_2016} \citeauthor{vazquez_maintaining_2016}
also assume a single conversation within an F-formation while developing a policy
for a robot to be aware of a single focus of attention of the conversation.

\hfill \break
\textbf{Estimating involvement.} In a conversation, the floor is typically held
by a single participant at a time \cite{sacks_simplest_1974}. What then
characterizes the silent participants in a conversation group? The following works
demonstrate that the task of estimating participant involvement is subjective
in nature, and that gaze behaviour and turn-taking patterns can be informative.

\citeauthor{zhang_beyond_2016} \cite{zhang_beyond_2016, zhang2018social} study
the task of detecting associates of an F-formation; members that are attached to
an F-formation but do not have full status \cite{kendon_conducting_1990}. They
argue that the labeling of conversation groups is not an objective task.
Collecting multiple annotations of perceived associates, they demonstrate how
detecting them can improve initial estimates of full-members of an F-formation.
Oertel, Funes Mora, Gustafson, and Odobez \cite{oertel_deciphering_2015}
characterize silent particiants into multiple categories (attentive listener,
side participant, bystander) from audiovisual cues. \citeauthor{oertel_gaze-based_2013}
\cite{oertel_gaze-based_2013} also show that it is possible to estimate individual
engagement and group involvement in a multiparty corpus by analysing the
participants' eye-gaze patterns. \citeauthor{bohus_managing_2014} \cite{bohus_managing_2014}
propose a self-supervised method for forecasting disengagement with an interactive robot
using a conservative heuristic. The heuristic is constructed by leveraging
features that capture how close the participant is, whether a participant is
stationary or moving, and whether a participant is attending to the robot.

Some works also used turn-taking features to estimate some notion of involvement.
\citeauthor{pentland_perception_2006} \cite{pentland_perception_2006} measured
engagement by the z-scored influence each person has on the other's turn-taking
for a pair of participants. \citeauthor{hung_estimating_2010} \cite{hung_estimating_2010}
found that the pause duration between an individual's turns, aggregated at group
level, is highly predictive of cohesion in small group meetings.

\hfill \break
\textbf{Schisming.} In a conversation with at least four participants, the
conversation sometimes splits up into two or more conversations. This
transformation is referred to as a \textit{schism} \cite{sacks_simplest_1974}
or \textit{schisming}. One of the earliest allusions to the phenomenon of
schisming based on anecdotal evidence occurs in the work of \citeauthor{goffman_behavior_1966},
who suggested that a gathering of two participants \textit{exhausts} an encounter
and forms a \textit{fully-focused gathering} \parencite[p. 91]{goffman_behavior_1966}.
With more than two participants, there may be persons officially present in the
situation who are not themselves so engaged. These \textit{bystanders} change
the gathering into a \textit{partly-focused} one. If more than three persons are
present, there may be more than one encounter carried on in the same situation,
resulting in a \textit{multifocused} gathering.

In subsequent work, \citeauthor{sacks_simplest_1974} \cite{sacks_simplest_1974}
and \citeauthor{goodwin_forgetfulness_1987} \cite{goodwin_forgetfulness_1987}
both indicated that the co-existence of two turn-taking systems is the most
decisive characteristic of schisming. This view was supported by \citeauthor{egbert_schisming:_1997},
who demonstrated that although schisming is a participation framework with two
simultaneous conversations, each with its own turn-taking system, there is an
interface between them during schisming \cite{egbert_schisming:_1997}. She also
makes a systematic differentiation between overlap and simultaneous talk during
schisming. In overlap, simultaneous speakers compete for the floor, an event
usually resolved by returning to \textit{one-speaker-at-a-time}. In schisming
by contrast, simultaneous speakers orient to one of two distinct floors, an event
which if resolved successfully, results in the establishment of two floors
\parencite[p. 43]{egbert_schisming:_1997}. Overlapping speech is therefore
expected to occur throughout the lifespan of all conversation floors within an
F-formation.

%% file: Sections/Methodology.tex
\section{Methodology} \label{sec:Methodology}

In this section we build upon the previously discussed concepts to propose
using simultaneous speakers in an F-formation as an initial conservative
indicator of the existence of distinct conversation floors.

A common concern with observing groups of conversing people is the potential
violation of privacy. In our experience with collecting group interaction datasets,
participants often regard having their microphone data recorded and transcribed
as being more invasive than being captured on video. In these situations, the
lack of verbal information makes it extremely challenging to infer the topics
being discussed. How can we then investigate the existence of distinct
conversations? Two observations could prove useful:

\hfill \break
\noindent \textbf{Inferring schisms without audio data.} The relationship between
body movements such as gestures and speech has been long established in literature
\cite{mcneill_language_2000}. Some works have shown promising results in
estimating the presence of voice activity from automated gestural analysis or
accelerometer data \cite{hung_speech/non-speech_2010, gedik_speaking_2016, gedik_personalised_2017}.
It therefore seems feasible that speaker turns can be automatically estimated
without audio data. Combined with the observation that the co-existence of two
turn-taking systems is the most decisive characteristic of schisming, we argue
that it is in turn reasonable to explore the inference of schisms without audio
data through speaking turns.

\hfill \break
\noindent \textbf{Linking schisming to floors and F-formations.} While \citeauthor{egbert_schisming:_1997}
does explicitly use the term \textit{floor} to describe the conversations
resulting from a schism, it is useful to observe how this relates back to
Edelsky's view of floors. Edelsky defined floors in terms of the acknowledged
\textit{what's-going-on} within a psychological time space. The object of focus here
could either be a topic or some other function. To borrow Goffman's terms, a schism
effectively changes a gathering into a \textit{multifocused} one, where each
object of focus can be viewed to correspond to a floor in Edelsky's definition.
However, if the participant's lower bodies remain configured such that their
transactional segments overlap to produce a common o-space, they would still
remain in the same F-formation even if the conversation has undergone a schism
into two or more distinct floors. Fig.~\ref{fig:concept} depicts this situation
conceptually.

\hfill \break
Combining these two broad observations, we argue that it is feasible to explore
the existence of distinct conversation floors within an F-formation without audio
data, whilst capturing speaker turns from visual observations. We propose to start
with the following metric. Given a sliding window $w$ of speaking duration $d$,
we consider a \textit{speaker} to be a participant who speaks for the entire
duration $d$. The number of simultaneous \textit{speakers} thus defined corresponds
to the number of distinct conversation floors at that position of $w$, since they
correspond to speaking turns in distinct floors.

\begin{figure}[b]
\makebox[\linewidth][c]{\includegraphics[width=\linewidth]{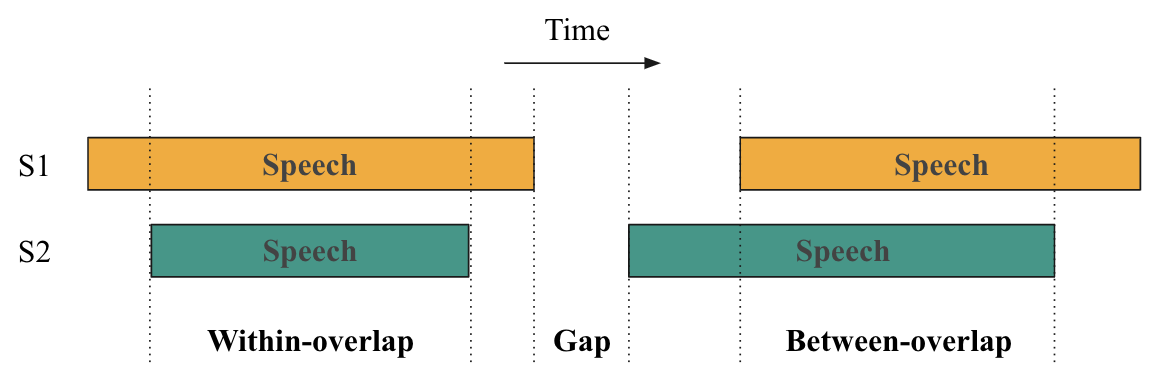}}
\caption{Illustration of gaps, within-overlaps, and between-overlaps for two
speakers (S1 and S2) within the same floor. The scheme was originally proposed
by \citeauthor{heldner_pauses_2010}\cite{heldner_pauses_2010} and adopted by
\citeauthor{levinson_timing_2015} in their analysis\cite{levinson_timing_2015}.}
\label{fig:overlaps}
\end{figure}

Of course, the metric is inextricably tied to the duration $d$ being considered;
too short a duration, and the concurrent turns might capture either backchannels
or the overlapping speech within the same floor as described in Egbert's work.
However, a reasonably long duration would capture the speaking turns of participants
holding distinct floors. This leads to the question: what qualifies
as a reasonable choice for $d$ to differentiate overlaps within a floor from
turns in distinct floors? In our experiments, we set the lower bound of $d$ at
one second. Here we provide evidence from literature to justify this choice.

\hfill \break
\noindent \textbf{Choice of speaking window duration.} In a study of gaps and
overlaps in conversations, \citeauthor{heldner_pauses_2010} report that on
average 40\% of the speaker transitions in their corpora involved
overlaps (including any overlap of over 10 ms) \cite{heldner_pauses_2010}. These
represent overlaps for competing for the floor. As for the duration of these
overlaps, their histogram makes clear that the duration follows a mode of 50 ms
in the Spoken Dutch Corpus, with a mean of 610 ms, and median of 470 ms, all under
one second. In a follow-up detailed statistical analysis, \citeauthor{levinson_timing_2015}
differentiate between types of overlaps: \textit{between-overlaps}, that refer
to overlaps where the floor was transfered without a silent gap between speakers;
and \textit{within-overlaps}, where overlapping speech occured in between a speaking
turn and did not result in a transfer of floor \cite{levinson_timing_2015}.
Fig.~\ref{fig:overlaps} illustrates these types of overlaps. They used the
Switchboard Corpus of English telephone conversations for their analysis,
and found that only 3.8\% of the signal corresponded to simultaneous speech of
both speakers. This fits well with Sacks and colleagues' observations that
``overwhelmingly, one party speakes at a time'' \parencite[p. 700]{sacks_simplest_1974},
for physically situated embodied social interactions. As for the duration,
\textit{between-overlaps} had a modal duration of 96 ms, a median of 205 ms,
a mean of 275 ms. On the other hand, \textit{within-overlaps} exhibited an
estimated modal duration of 350 ms, a median of 389 ms, a mean of 447 ms.
Further, of all the overlaps annotated, 73\% involved a backchannel. These
statistics indicate that choosing a lower bound for $d$ would reasonably capture
simultaneous speech that does not belong to the same floor.

As for the upper bound, a reasonable value should be at least greater than the
average turn duration of a speaker. Using the same operationalization proposed
in \cite{heldner_pauses_2010}, \citeauthor{levinson_timing_2015} report that
contiguous speech delimited by a silent interval of at least 180 ms had
a mean duration of 1680 ms, and a median of 1227 ms.

%% file: Sections/Dataset.tex
\section{Dataset} \label{sec:Dataset}

\begin{figure}[t]
\makebox[\linewidth][c]{\includegraphics[width=\linewidth]{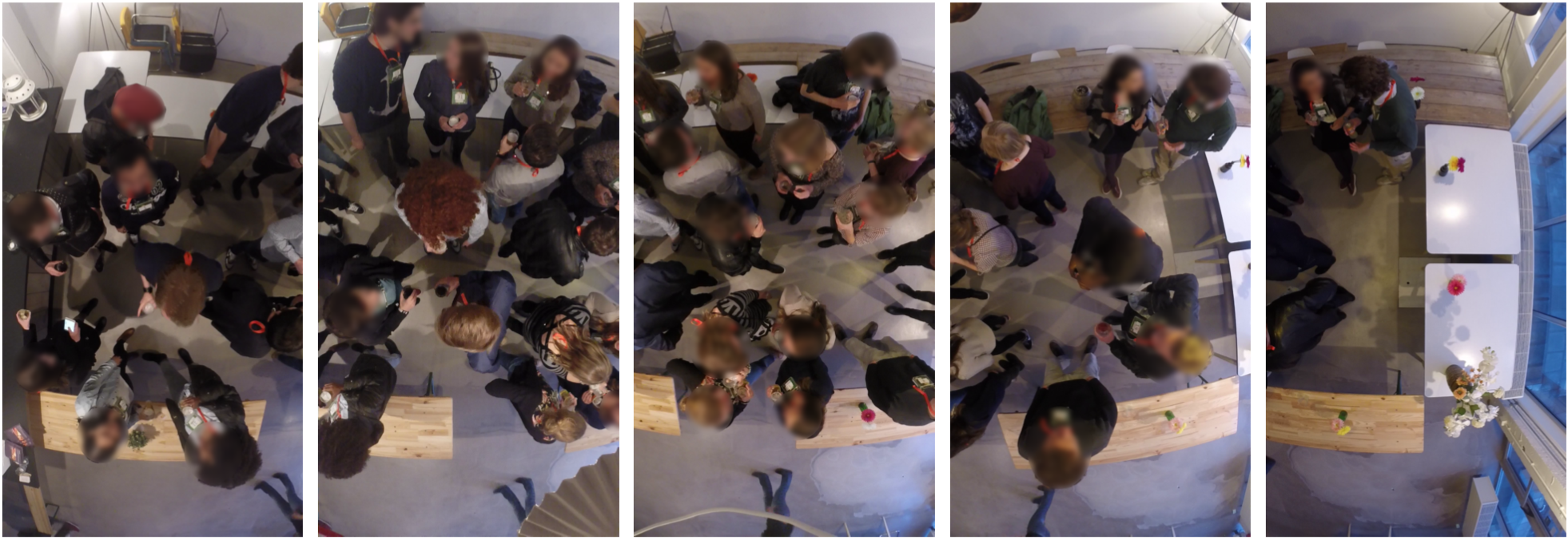}}
\caption{Snapshots of the mingling session (Cameras 1-5) in MatchNMingle.}
\label{fig:MnM}
\end{figure}

For this study, we use the publicly available \textit{MatchNMingle} dataset \cite{cabrera-quiros_matchnmingle_2018} that records in-the-wild interactions of 92 people during speed-dates followed by a cocktail party. Three sessions of speed-dates and mingling were recorded in all across three days. We specifically focus on the cocktail party recordings that capture free standing conversations between participants. Fig.\ref{fig:MnM} shows the video recordings from five cameras on the last day of data collection. The participants were not given a script to follow and were free to choose the participants they wished to interact with. This allows us to study naturally evolving F-formations and conversation floors in an in-the-wild setting.

\hfill \break
\noindent \textbf{Dataset Statistics.} The dataset consists of a total of 92 single, heterosexual participants (46 women: 19-27 years with a mean age of 21.6 years and standard deviation of 1.9 years; and 46 men: 18-30 years with a mean age of 22.6 years and standard deviation of 2.6 years). Over 45 minutes of free mingling interaction were recorded for each of the three days; 56 minutes on the first, 50 minutes on the second, and 45 minutes on the third, respectively.

\hfill \break
\noindent \textbf{Annotations.} The dataset provides of annotations for both F-formations and a variety of social actions. The F-formations were annotated directly from a video of the interacting participants captured from overhead cameras. The annotations were made for every second for an interval of 10 minutes per day. Each F-formation annotation provides the participant IDs for its members and the start and end times delimiting the lifetime of the F-formation. In all, 174 F-formations were annotated across 30 minutes. Of these, we filtered out those with cardinality less than four, and those for which a participant was found to leave the field of view of the cameras. This left us with 34 F-formations for our experiments.

Of the social actions annotated, we only use the Speaking Status---defined as whether or not a person is speaking. The social actions were annotated for a 30 minute segment for each day, by eight annotators hired for the task and trained by an expert. The annotations were made at the frame level using a tool that allowed for interpolation across frames. In all, 20 annotations per second for each social action are provided. Further, the speaking status is estimated from video alone, by observing lip movements or inferring from the participants' head and body gestures.

%% file: Sections/Experiments.tex
\section{Experiments} \label{sec:Experiments}

\begin{figure*}
\centerline{\includegraphics[width=\textwidth]{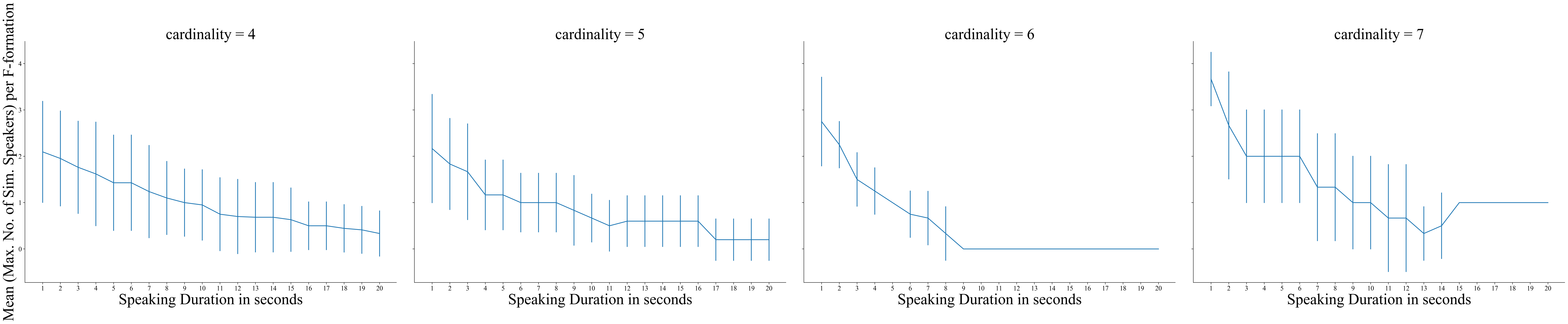}}
\caption{Plotting the effect of varying the speaking duration threshold $d$ on
the number of simultaneous speakers per cardinality of F-formation. To aggreagate
the data from each F-formation, the maximum of the number of simultaneous speakers
is considered over all the sliding window positions across the lifetime of the
F-formation. The y-axis plots the Mean (Maximum number of simultaneous speakers
over window positions) over F-formations.}
\label{fig:nfloors}
\end{figure*}

We perform two sets of experiments: first we identify the number of simultaneous
speakers in an F-formation using the methodology described in
Section~\ref{sec:Methodology}, and then evaluate whether the number of members
in an F-formation (cardinality) affects the speaking duration of simultaneous
speakers.

\hfill \break
\noindent \textbf{Simultaneous Speakers in an F-formation.} The purpose of this
experiment is to evaluate the following---can we infer the existence of distinct
conversation floors within an F-formation from simultaneous speaker turns?
To recap, this intuition build upon early work on schisming indicating that the
co-existence of two turn-taking systems is the most decisive characteristic of
distinct conversation floors \cite{sacks_simplest_1974, goodwin_forgetfulness_1987}.
Here we consider F-foramtions of cardinality four and above, since the possibility
of distinct conversations occurs only for those F-formations.

We slide a window $w$ of duration $d$ across the lifetime of the F-formation in
steps of one second. For every position of $w$, we count the number of participants
with a positive speaking status for the entire duration $d$. We plot the maximum
number of simultaneous speakers over all positions of $w$. Following the formulation
described in Section~\ref{sec:Methodology}, this represents the maximum number
of distinct conversation floors that were observed during the life-time of the
F-formation. We vary $d$ from 1-20 seconds to guard against the possibility
that the smaller values of $d$ might capture co-narration or overlaps within
the same floor. The upper bound of 20 seconds was chosen as sanity check; we
expected to see very few speakers have a speaking turn that long.

The $max$ operator was chosen to aggregate the number of simultaneous speakers
across all window positions into the most conservative measure for what this
experiment seeks to evaluate. A value of one for the maximum number speakers
over all positions of $w$ would indicate that only a single conversation floor
existed within the F-formation. Therefore, observing values greater than one
for the $max$ metric would indicate the presence of distinct floors with more
certainty than other choices of summarizing statistics.

Fig.~\ref{fig:nfloors} plots the mean number of distinct conversation floors
per F-formation against varying values of $d$, per cardinality of F-formation.
Cardinality here refers to the number of members in an F-formation. As a sanity
check, we would expect the numbers upper-bounded by the number of people in the
F-formation; at worst, every person in the F-formation speaks simultaneously
to compete for the floor they are a part of. On the same note, we observe that
the starting mean values all seem reasonable: about 2 for cardinalities four and
five, about 3 for cardinality six, and about 4 for cardinality seven. Assuming
that it is common for speakers to have at least one conversing partner, we would
expect about half the number of simultaneous speakers as members in an F-formation.
Our minimum choice of $d$ was chosen to be greater than the modal duration of
overlaps found in previous work \cite{levinson_timing_2015}, so it is less likely
that the lower turn durations capture competing overlaps for the same floor.
Moreover, at the average turn length of about two seconds observed by
\citeauthor{levinson_timing_2015} \cite{levinson_timing_2015}, we observe that the
maximum number of simultaneous speakers is greater than one at all cardinalities
considered. This suggests that the simplifying assumption from previous research
of a single conversation within an F-formation is insufficient.

We also observe a decreasing trend for the curves in Fig.~\ref{fig:nfloors}. This
seems intuitive, as it is much less likely that participants would speak
for the entire duration of a window as $d$ increases. Interestingly, there is a
single example of a speaker speaking for 20 seconds in an F-formation of cardinality seven.
On closer inspection, this turned out to be an error in speaking status annotation,
and we manually fixed this error for subsequent analysis.

\hfill \break
\noindent \textbf{Effect of cardinality on turn duration of simultaneous speakers.}
\citeauthor{sacks_simplest_1974} observed that there is a
``pressure for minimization of turn size, distinctively operative with three or
more parties'' \parencite[p. 713]{sacks_simplest_1974}. They note that the
possibility of a schism introduced by the fourth participant may influence the
turn-taking system by `spreading the turns around' if there is an interest in
retaining participants in the conversation. However, they concede that this
effect is equivocal, since turn distribution can also be used for encouraging
schisming. In this experiment, we explore this effect and pose the question as
follows: for a given speaking turn duration $d$, do we observe a decrease in
the maximum number of conversation floors observed over an F-formation's lifetime
with an increase in the cardinality of an F-formation?

\begin{table}[b]
\caption{Generalized Linear Model Regression Results}
\begin{center}
    \begin{tabular}{lcccc}
    \hline
                                        & \textbf{Coef ($\boldsymbol{\beta}$)} & \textbf{Std Err} & \textbf{z} & \textbf{P$>$$|$z$|$} \\
    \hline
    \textbf{Intercept}                  &        0.0626        &        0.339     &     0.184  &         0.854        \\
    \textbf{Turn-duration}              &        0.0057        &        0.002     &     2.296  &         0.022        \\
    \textbf{Cardinality}                &        0.1869        &        0.072     &     2.603  &         0.009        \\
    \textbf{Turn-duration:Cardinality}  &       -0.0025        &        0.001     &    -4.543  &        0.000006      \\
    \hline
    \end{tabular}
\label{tab:glm}
\end{center}
\end{table}

Qualitatively, this corresponds to the steepness of fall-off of the curves in
Fig.~\ref{fig:nfloors}. It seems that the the curves for
cardinality six and seven falloff more steeply than those for
cardinalities four and five. To quantitatively test if cardinality has an
effect, we fit a Generalized Linear Model (GLM) to the same data as in the
previous experiment with an interaction factor between cardinality and the speaking turn
duration $d$. Specifically, we assume the maximum number of simultaneous
speakers observed over the lifetime of each F-formation, $y_i$ to be realizations of
independent Poisson random variables, with \( Y_i \sim P(\mu_i) \) and model
$\mu_i$ as follows:

\begin{equation}
    \log(\mu_i) = \beta_0 + \beta_1*d_i + \beta_2*c_i + \beta_3*d_i*c_i
\end{equation}

where $d_i$ refers to the duration of the speaking window, and $c_i$ refers to
the cardinality for the $i$th observation. The $\beta$s refer to the regression
coefficients. The GLM was fit using the \textit{statsmodels} python package.
The results of the GLM regression test are provided in Table~\ref{tab:glm}.
We conclude that cardinality and the two-way interaction between cardinality and
turn duration are statistically significant at a significance level of 0.01.
Turn duration is itself significant at a significance level of 0.05.

While the previous test tells us that turn duration and cardinality are significant,
we still need to perform post-hoc comparisons to ascertain the differences between
the cardinalities. We fit multiple GLMs to each possible pair
of cardinalities being considered and correct the corresponding p-values using
the Bonferroni correction for multipe testing. Table~\ref{tab:glm-posthoc} provides
the corrected p-values for the post-hoc comparisons. From the last column, we
find that cardinality and its interaction with turn-duration are significant
between the cardinalities \{4, 6\}, and \{5, 6\} at a significance level of 0.001.

One potential limitation of this analysis is the imbalance in the number of
F-formations of different cardinalities. F-formations of cardinality four were
the most common in the data, with reasonable number of samples to infer a pattern.
We believe that the intuition of cardinality and its interaction with speaking turn duration
being significant is still a sound intuition, although the statistical significance
should perhaps be viewed within the context of the number of F-formations we see in the data.
Fig.~\ref{fig:counts} plots the number of observations that contributed to the
graphs in Fig.~\ref{fig:nfloors}.

\begin{table}[t]
\caption{Nominal \textit{P}-values for Six Post-Hoc GLM Regression Comparisons}
\begin{center}
    \begin{tabular}{|c|cccc|}
    \hline
    \textbf{Cardinality Pairs} & \textbf{Intercept ($\boldsymbol{\beta_0}$)} & \textbf{d ($\boldsymbol{\beta_1}$)} & \textbf{c ($\boldsymbol{\beta_2}$)} & \textbf{d:c ($\boldsymbol{\beta_3}$)} \\
    \hline
    \textbf{4-5}               &         0.196                &         0.855         &      0.794           &           0.403        \\
    \textbf{4-6}               &         0.364                &         0.0007        &      0.010           &           0.00002\textsuperscript{*}  \\
    \textbf{4-7}               &         0.697                &         0.428         &      0.030           &           0.009        \\
    \textbf{5-6}               &         0.079                &         0.0008        &      0.016           &           0.00016\textsuperscript{*}  \\
    \textbf{5-7}               &         0.434                &         0.413         &      0.043           &           0.052        \\
    \textbf{6-7}               &         0.275                &         0.006         &      0.657           &           0.024        \\
    \hline
    \end{tabular}
    \begin{tablenotes}
        \small
        \item d = turn-duration, c = cardinality, d:c = interaction-factor.
        $\beta$s denote the corresponding regression coefficients. *
        denotes significance at a threshold of $0.001$ after Bonferroni correction for six tests.
      \end{tablenotes}
\label{tab:glm-posthoc}
\end{center}
\end{table}

%% file: Sections/Conclusion.tex
\section{Conclusion} \label{sec:Conclusion}

In this study, we presented an initial exploration into unifying the spatial
and temporal perspectives of a free-standing conversing group. Specifically, we
proposed using simultaneous speaking turns as an indicator for the existence of
distinct conversation floors. In the absence of audio data to identify the topics
being discussed, our proposed metric can be used to gain a deeper understanding
of the conversation dynamics within an F-formation, since speaking turns can be
inferred from visual or wearable-sensor data. Our experiments demonstrate that
at an average turn duration of two seconds for humans \cite{levinson_timing_2015},
there is evidence of multiple conversation floors within a single F-formation.
Further, we found that an increase in cardinality of an F-formation correlates
with a decrease in turn duration of simultaneous speakers, specifically between
F-formations of sizes \{4,6\}, and \{5,6\} in our data. A deeper analysis would
be required to identify whether the differences in F-formations of cardinality six
hold across datasets, with preferably more examples of F-fomrations of size six
and greater. In this initial approach to the problem, our study does not account
for the behaviour of the silent participants, or the evolution of turn taking
dynamics within a floor. These remain promising avenues to explore for future works.

\begin{figure}[t]
\makebox[\linewidth][c]{\includegraphics[width=\linewidth]{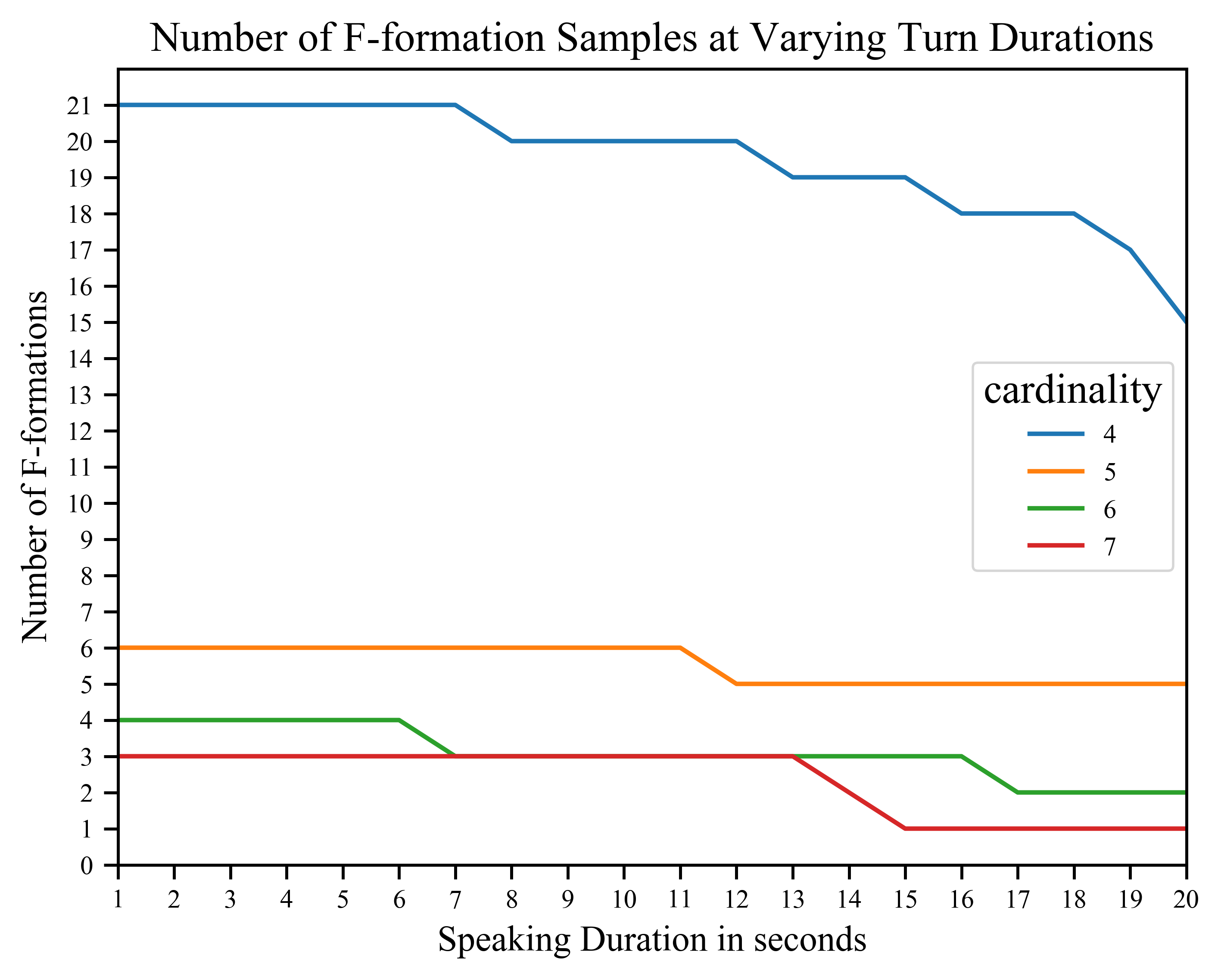}}
\caption{Number of F-formations at different speaking turn durations.}
\label{fig:counts}
\end{figure}